\documentclass[11pt]{article}
\usepackage{moriond,epsfig}

\def\Journal#1#2#3#4{{#1} {\bf #2}, #3 (#4)}

\def\NPB{{\em Nucl.\ Phys.} B}
\def\PLB{{\em Phys.\ Lett.}  B}
\def\PRL{\em Phys.\ Rev.\ Lett.}

\def\JHEP{{\em J.\ High Energy Phys.}}
\def\APJ{{\em Astrophys.\ J.}}
\def\APJS{{\em Astrophys.\ J.\ Suppl.}}

\def\be{\begin{equation}}
\def\ee{\end{equation}}
\def\bea{\begin{eqnarray}}
\def\eea{\end{eqnarray}}

\newcommand{\befig}{\begin{figure}}
\newcommand{\efig}{\end{figure}}
\newcommand{\gtsim}{\;\lower-0.45ex\hbox{$>$}\kern-0.77em\lower0.55ex\hbox{$\sim$}\;}
\newcommand{\ltsim}{\;\lower-0.45ex\hbox{$<$}\kern-0.77em\lower0.55ex\hbox{$\sim$}\;}
\newcommand{\gl}{\ensuremath{\tilde{g}}}
\newcommand{\ax}{\ensuremath{\tilde{a}}}

\newcommand{\keV}{\mbox{keV}}
\newcommand{\MeV}{\mbox{MeV}}
\newcommand{\GeV}{\mbox{GeV}}

\newcommand{\MPl}{\mathrm{M}_{\mathrm{Pl}}}
\newcommand{\equil}{\mathrm{eq}}

\begin{document}
\vspace*{4cm}
\title{AXINOS AS DARK MATTER IN THE UNIVERSE~\footnote{Talk presented by F.D.S.\ 
at the 39th Rencontres de Moriond,
La Thuile, Italy, March 28 - April 4, 2004.} }

\author{ ARND~BRANDENBURG and FRANK~D.~STEFFEN }

\address{DESY Theory Group, Notkestrasse 85, D-22603 Hamburg, Germany}

\maketitle\abstracts{The axino is the fermionic superpartner of the
  axion. Assuming the axino is the lightest supersymmetric particle
  and stable due to $R$-parity conservation, we compute the relic
  axino density from thermal reactions in the early Universe.  From
  the comparison with the WMAP results, we find that thermally
  produced axinos could provide the dominant part of cold dark matter,
  for example, for an axino mass of 100~keV and a reheating
  temperature of $10^6$~GeV.}

\section{Introduction}

If supersymmetry is realized in nature and the strong CP problem is
solved by the Peccei-Quinn (PQ) mechanism, the axino $\ax$ should
exist as the fermionic superpartner of the
axion.\cite{Nilles:1981py+X} As the interactions of the electrically
and color neutral axion supermultiplet are suppressed by the PQ~scale
$f_a/N \gtsim 5\times 10^{9}\,\GeV$, the axino couples only very
weakly to the MSSM particles. If the axino is the lightest
supersymmetric particle (LSP) and if $R$-parity is conserved, axinos
will be stable and exist as dark matter in the Universe. Then,
depending on the axino mass $m_{\ax}$ and the temperature $T$ of the
primordial plasma, axinos could play an important role in the cosmos.
In particular, they could provide the dominant part of cold dark
matter.

In this talk we consider the thermal production of stable axinos in
the early Universe. Assuming that inflation has governed the earliest
moments, any initial population of axinos has been diluted away by the
exponential expansion during the slow-roll phase.  The thermal
production of axinos starts then after completion of the reheating
phase at the reheating temperature~$T_R$.  We restrict our
investigation to $f_a/N \gtsim T_R \gtsim 10^4\,\GeV$, where the
U(1)$_{\mathrm{PQ}}$ symmetry is broken and axino production from
decays of particles out of equilibrium is
negligible.\cite{Covi:2001nw} Based on our computation of the thermal
axino production rate within thermal field
theory,\cite{Brandenburg:2004du} we discuss the possibility of stable
axinos forming the dominant component of cold dark matter.  The
results presented are extracted from Ref.~3 where more details can be
found.

\section{Relic Axino Abundance from Thermal Production in the Early Universe}

We concentrate on the axino-gluino-gluon interactions given by the
dimension-5 interaction term
\be
        {\cal L}_{\ax\gl g}= 
        i\,\frac{g^2}{64\pi^2 (f_a/N)}\,\bar{\ax}\,\gamma_5
        \left[\gamma^{\mu},\gamma^{\nu}\right]\,\gl^a\, G^a_{\mu\nu}
        \ ,
\label{Eq:L_agG}
\ee
with the strong coupling $g$, the gluon field strength tensor
$G^a_{\mu\nu}$, the gluino $\gl$, and $N$ being the number of quarks
with PQ~charge. These interactions govern the thermal axino production
at high temperatures.  For $f_a/N = 10^{11}\,\GeV$, Rajagopal,
Wilczek, and Turner have estimated an axino decoupling temperature of
$T_D\approx 10^9\,\GeV$ by comparing the corresponding axino
interaction rate with the Hubble parameter $H$ for the early
radiation-dominated epoch.\cite{Rajagopal:1990yx}

For $T_R \gtsim T_D$, axinos have been in thermal equilibrium with the
primordial plasma. Their abundance is then given by the equilibrium
number density of a relativistic Majorana fermion while contributions
from axino production and annihilation processes at smaller
temperatures are negligible. Thus, the axino yield and the axino
density parameter are given respectively by
\bea
        Y_{\ax}^{\equil}  
        & = & \frac{n_{\ax}^{\equil}(T_D)}{s(T_D)}
        \approx 1.8 \times 10^{-3} \ ,
\label{Eq:Y_equil} \\
        \Omega_{\ax}^{\equil}h^2
        & = & m_{\ax}Y_{\ax}^{\equil} s(T_0)h^2/\rho_c
        \approx\frac{m_{\ax}}{2\,\keV}
\label{Eq:Omega_axino_equil} \ ,
\eea
where $n_{\ax}$ is the axino number density,
$s(T)=2\pi^2g_{*S}(T)T^3/45$ is the entropy density, $h$ parameterizes
the Hubble constant $H_0 = 100\,h\,\mbox{km/sec/Mpc}$, $T_0$ is the
present temperature of the cosmic photon background radiation, and
$\rho_c$ is the critical density so that $\rho_c/[s(T_0)h^2]=3.6\times
10^{-9}\,\GeV$. In $s(T_D)$ we have used the MSSM value
$g_{*S}(T_D)=228.75$ for the number of effectively massless degrees of
freedom being in thermal equilibrium with the primordial plasma at
$T_D$.

For $T_R < T_D$, axinos have not been in thermal equilibrium after
inflation. Here the evolution of $n_{\ax}$ with cosmic time $t$ can be
described by the Boltzmann equation with a collision term $C_{\ax}$
accounting for both the axino production and disappearance in thermal
reactions in the primordial plasma,
\be
    \frac{dn_{\ax}}{dt} + 3 H n_{\ax} = C_{\ax}
    \ .
\label{Eq:Boltzmann}
\ee
The disappearance processes can be neglected for $T_R$ sufficiently
below $T_D$. The collision term then is given by integrating the
thermal axino production rate. We have computed this rate for axinos
with $E\gtsim T$ in SUSY QCD using the Braaten-Yuan prescription and
hard thermal loop (HTL) resummation.\cite{Brandenburg:2004du} A finite
result is obtained in a gauge-invariant way, which takes into account
Debye screening in the primordial plasma and does not depend on any
{\em ad hoc} cutoffs. By numerical integration of our result for the
axino production rate, we obtain the collision term
\be
        C_{\ax}(T) \Big|_{T_R<T_D}\!\!
        \approx
        {(N_c^2 - 1) \over (f_a/N)^2} 
        {3\zeta(3) g^6 T^6 \over 4096\pi^7}
        \Bigg[\ln\left(\frac{1.380\,T^2}{m_g^2}\right)(N_c+n_f) 
        + 0.4336\,n_f \Bigg]
        \ ,
\label{Eq:C_axino}
\ee
where $N_c = 3$ is the number of colors, $n_f = 6$ is the number of
color triplet and anti-triplet chiral multiplets, $\zeta(3)\approx
1.2021$, and $m_g=gT\sqrt{(N_c+n_f)/6}$ is the thermal gluon mass in
SUSY QCD. Assuming conservation of entropy per comoving volume, the
Boltzmann equation can be integrated analytically. The resulting axino
yield and axino density parameter read respectively
\bea
        Y_{\ax}(T_0)
        &\!\!\approx\!\!\!& \frac{C_{\ax}(T_R)}{s(T_R) H(T_R)}
        = 2.0\times 10^{-7}\,  g^6 \ln \left(\frac{1.108}{g}\right)\!
        \left(\frac{10^{11}\,\GeV}{f_a/N}\right)^{\! 2}\!\!
        \left(\frac{T_R}{10^4\,\GeV}\right)
        \ ,
\label{Eq:axino_yield}\\
        \Omega_{\ax}h^2 
        &\!\!=\!\!\!&  m_{\ax}Y_{\ax}(T_0) s(T_0)h^2/\rho_c
        = 5.5\,g^6 \ln\left( \frac{1.108}{g}\right) 
        \bigg(\frac{m_{\ax}}{0.1\,\GeV}\bigg)\!
        \left(\frac{10^{11}\,\GeV}{f_a/N}\right)^{\! 2}\!\!
        \left(\frac{T_R}{10^4\,\GeV}\right),\quad
\label{Eq:Omegah2_axino}
\eea
where we have used the Hubble parameter describing the
radiation-dominated epoch $H(T)=\sqrt{g_*(T)\pi^2/90}\, T^2/\MPl$ with
$\MPl \approx 2.4\times 10^{18}\,\GeV$ and
$g_*(T_R)=g_{*S}(T_R)=228.75$.

\pagebreak

In Fig.~\ref{Fig:axino_density_and_DM_constraints}a, our result for
the relic axino density $\Omega_{\ax}h^2$ is illustrated as a function
of the reheating temperature $T_R$ for a PQ~scale of $f_a/N =
10^{11}\,\GeV$ and an axino mass of $m_{\ax} = 1\,\keV$ (dotted line),
$100\,\keV$ (dashed line), and $10\,\MeV$ (solid line).  The 1-loop
running of the strong coupling in the MSSM is taken into account by
replacing $g$ with
$g(T_R)=[g^{-2}(M_Z)\!+\!3\ln(T_R/M_Z)/(8\pi^2)]^{-1/2}$, where the
value of the strong coupling at the Z-boson mass $M_Z$,
$g^2(M_Z)/(4\pi)=0.118$, is used as input. For $T_R$ above $T_D
\approx 10^9\,\GeV$, the relic axino
density~(\ref{Eq:Omega_axino_equil}) is independent of $T_R$ as shown
for $m_{\ax} = 1\,\keV$ by the horizontal line. There will be a smooth
transition instead of a kink once the axino disappearance processes
are taken into account.  The grey band in
Fig.~\ref{Fig:axino_density_and_DM_constraints}a indicates the WMAP
result\,\cite{Spergel:2003cb} on the cold dark matter density
(2$\sigma$ error) $\Omega_{\mathrm{CDM}}^{\mathrm{WMAP}} h^2 =
0.113^{+0.016}_{-0.018}$.

\section{Can Axinos Provide the Dominant Component of Cold Dark Matter?}

Assuming the axino is the LSP and stable due to $R$-parity
conservation, the first crucial criterion is whether the relic axino
abundance matches the observed abundance of cold dark matter. As one
can see from Fig.~\ref{Fig:axino_density_and_DM_constraints}a, there
are certain combinations of $m_{\ax}$ and $T_R$ for which the relic
axino density $\Omega_{\ax}h^2$ indeed agrees with the WMAP result on
the abundance of cold dark matter.
\befig[t] 
\epsfig{figure=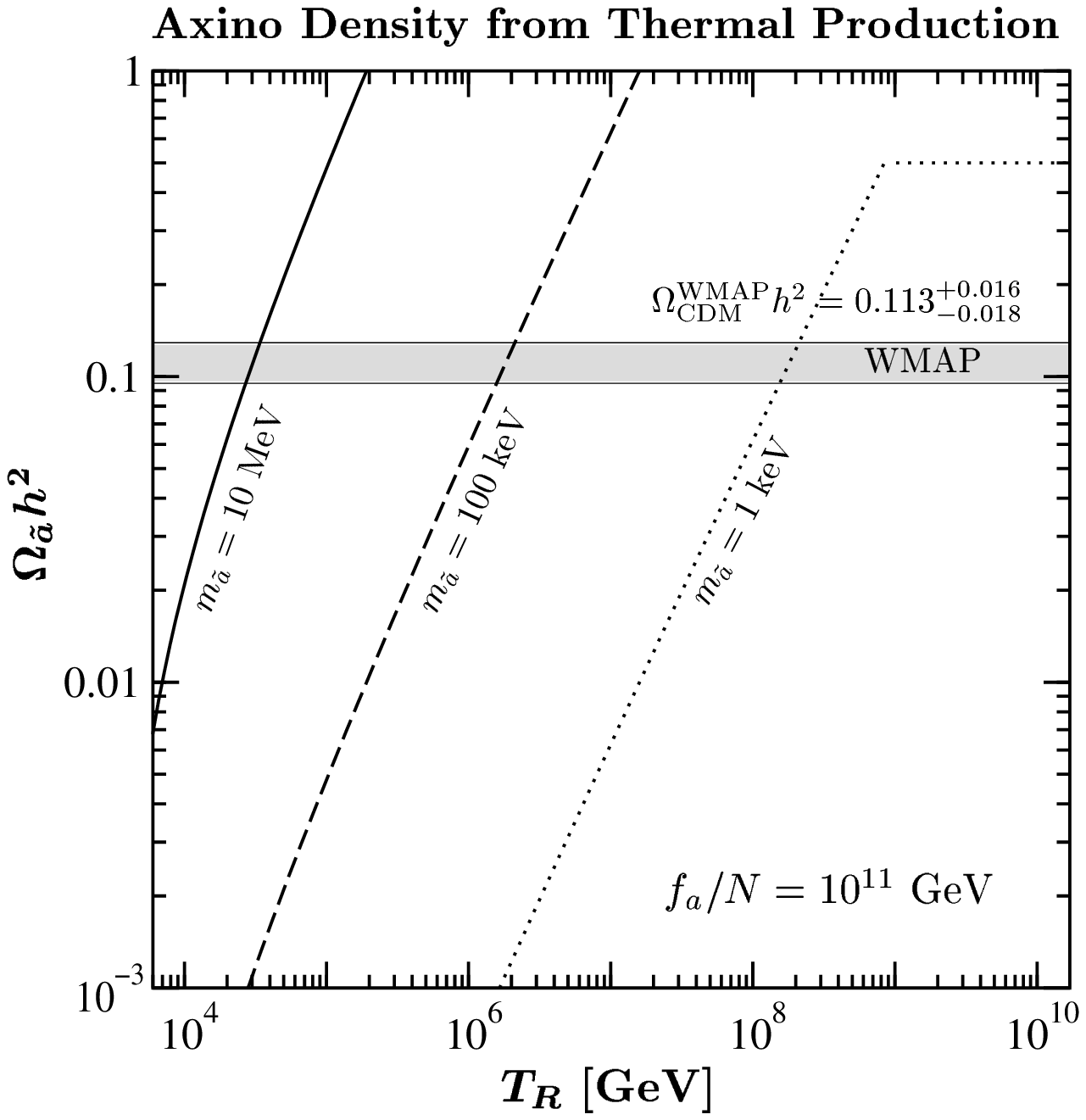,width=7.6cm}
\vskip -7.95cm
\hfill
\epsfig{figure=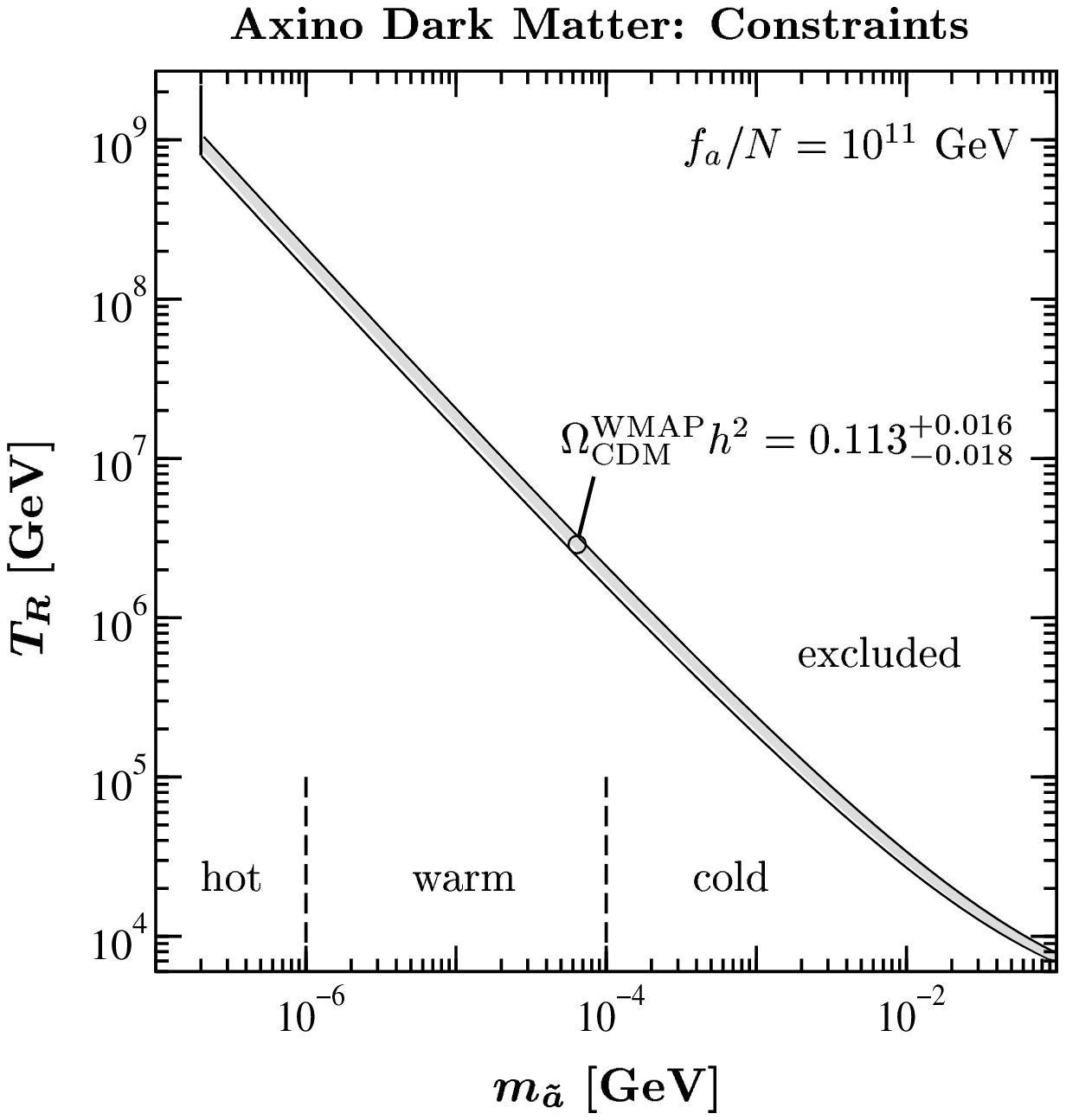,width=7.7cm}
\caption{
  (a) The axino density parameter $\Omega_{\ax}h^2$ as a function of
  the reheating temperature $T_R$ for a PQ~scale of $f_a/N =
  10^{11}\,\GeV$ and an axino mass of $m_{\ax} = 1\,\keV$ (dotted
  line), $100\,\keV$ (dashed line), and $10\,\MeV$ (solid line). (b)
  Constraints for axino dark matter from thermal production in the
  early Universe. For the combinations of $m_{\ax}$ and $T_R$ within
  grey band, the density parameter of thermally produced axinos
  $\Omega_{\ax}h^2$ (obtained for $f_a/N = 10^{11}\,\GeV$) agrees with
  the WMAP result on the cold dark matter density ($2\sigma$ errors)
  $\Omega_{\mathrm{CDM}}^{\mathrm{WMAP}} h^2 =
  0.113^{+0.016}_{-0.018}$.}
\label{Fig:axino_density_and_DM_constraints}
\efig

The next question is whether axino dark matter from thermal production
is sufficiently cold (i.e.\ non-relativistic sufficiently long before
matter-radiation equality) in order to explain the formation and power
spectrum of large-scale structures and the early reionization observed
by the WMAP satellite. As the thermal axino production rate shows
basically a thermal spectrum,\cite{Brandenburg:2004du} the dark matter
classification for particles with thermal spectrum can be adopted. For
an axino mass in the range $m_{\ax} \ltsim 1\,\keV$, $1\,\keV \ltsim
m_{\ax} \ltsim 100\,\keV$, and $m_{\ax} \gtsim 100\,\keV$, we refer to
hot, warm, and cold axino dark matter, respectively. Thus, axinos
could provide the dominant part of {\em cold} dark matter for
$T_R\ltsim 10^6\,\GeV$;
cf.~Fig.~\ref{Fig:axino_density_and_DM_constraints}. While our
classification is only an approximate guideline, simulations of early
structure formation show that dark matter of particles with mass $m_x
\ltsim 10\,\keV$ cannot explain the early reionization observed by the
WMAP satellite.\cite{Yoshida:2003rm}

Another important issue is whether the abundance of the primordial
light elements (D, $^3$He, $^4$He, Li) remains unaffected by the
production of axinos. The increase of the energy density at the time
of primordial nucleosynthesis due to the presence of the thermally
produced axinos is small and consistent with the observed abundance of
the light elements.\cite{Brandenburg:2004du} More model dependent and
possibly more severe are the nucleosynthesis constraints concerning
the potential destruction of the light primordial elements by photons
or hadronic showers from late decays of the NLSP into
axinos.\cite{Covi:1999ty,Covi:2001nw,Covi:2004rb} Anyhow, the axino
LSP is far less problematic than the gravitino LSP as the axino
interactions are not as strongly suppressed as the gravitino
interactions.

\section{Conclusion}

The relic axino density (obtained for $f_a/N = 10^{11}\,\GeV$) agrees
with the WMAP result ($2\sigma$ errors) on the cold dark matter
density for the $m_{\ax}$--$T_R$~combinations within the grey band
shown in Fig.~\ref{Fig:axino_density_and_DM_constraints}b and, in
particular, for $m_{\ax}=100\,\keV$ and $T_R \approx 10^6\,\GeV$.
Although relatively light for being cold dark matter, axinos with
$m_{\ax}=100\,\keV$ could still explain large-scale-structure
formation, the corresponding power spectrum, and the early
reionization observed by WMAP. For larger $m_{\ax}$, the matching with
the WMAP result does restrict $T_R$ to lower values.  Already $T_R
\approx 10^6\,\GeV$ is relatively small and excludes some models for
inflation and baryogenesis such as thermal
leptogenesis.\cite{Fukugita:1986hr+X} Note that the limits on $T_R$
shown in Fig.~\ref{Fig:axino_density_and_DM_constraints}b are relaxed
by one order of magnitude with respect to the ones found in Ref.~2.
This difference results from our computation of the thermal production
rate,\cite{Brandenburg:2004du} in which we have used the Braaten-Yuan
prescription with the HTL-resummed gluon propagator as opposed to the
pragmatic cutoff procedure used in Ref.~2. There is no upper limit on
$T_R$ for $m_{\ax}\ltsim 0.2\,\keV$. As the axinos have been in
thermal equilibrium with the primordial plasma for $T_R \gtsim
10^9\,\GeV$, the resulting relic axino
density~(\ref{Eq:Omega_axino_equil}) is completely determined by
$m_{\ax}$ and independent of $T_R$. The agreement with the WMAP result
is then achieved for any $T_R \gtsim 10^9\,\GeV$ as long as $m_{\ax}
\approx 0.2\,\keV$, which is the updated version of the
Rajagopal-Turner-Wilczek bound.\cite{Rajagopal:1990yx,Covi:2001nw}
However, axinos with $m_{\ax}\ltsim 0.2\,\keV$ are too light for being
cold dark matter. As warm or hot dark matter, they alone cannot
explain the formation and power-spectrum of large-scale structures
and, in particular, the early reionization observed by WMAP.
Nevertheless, even such a light axino can have important cosmological
implications provided it is the LSP. By destabilizing the other
$R$-odd particles including the gravitino, the light axino provides a
solution to the gravitino problem so that a reheating temperature of
$T_R\gtsim 10^{9}\,\GeV$ is no longer problematic.\cite{Asaka:2000ew}
Thus, with the light axino being the LSP, thermal
leptogenesis~\cite{Fukugita:1986hr+X} can coexist with SUSY and
explain the baryon asymmetry in the Universe. The cold dark matter
needed to explain structure formation could then be provided by the
axion.


\section*{References}
\begin{thebibliography}{99}

%
\bibitem{Nilles:1981py+X}
%
H.P.~Nilles and S.~Raby,
\Journal{\NPB}{198}{102}{1982};\\
%
J.E.~Kim and H.P.~Nilles,
\Journal{\PLB}{138}{150}{1984}.

%
\bibitem{Covi:2001nw}
L.~Covi, H.B.~Kim, J.E.~Kim, and L.~Roszkowski,
\Journal{\JHEP}{0105}{033}{2001}.

%
\bibitem{Brandenburg:2004du}
A.~Brandenburg and F.D.~Steffen,
arXiv: hep-ph/0405158.

%
\bibitem{Rajagopal:1990yx}
K.~Rajagopal, M.S.~Turner, and F.~Wilczek,
\Journal{\NPB}{358}{447}{1991}.

%
\bibitem{Spergel:2003cb}
D.N.~Spergel {\it et al.},
\Journal{\APJS}{148}{175}{2003}.

%
\bibitem{Yoshida:2003rm}
N.~Yoshida, A.~Sokasian, L.~Hernquist, and V.~Springel,
\Journal{\APJ}{591}{L1}{2003}.

%
\bibitem{Covi:1999ty}
L.~Covi, J.E.~Kim, and L.~Roszkowski,
\Journal{\PRL}{82}{4180}{1999}.
%
\bibitem{Covi:2004rb}
L.~Covi, L.~Roszkowski, R.~Ruiz de Austri, and M.~Small,
arXiv:hep-ph/0402240.

%
\bibitem{Fukugita:1986hr+X}
%
M.~Fukugita and T.~Yanagida,
\Journal{\PLB}{174}{45}{1986};\\
%
W.~Buchm\"uller, P.~Di Bari, and M.~Pl\"umacher,
arXiv:hep-ph/0401240.

%
\bibitem{Asaka:2000ew}
T.~Asaka and T.~Yanagida,
\Journal{\PLB}{494}{297}{2000}.

\end{thebibliography}
\end{document}